\documentstyle[psfig]{aa}

\newcommand{\h}{h_{50}}
\newcommand{\me}{m_{\rm e}}
\newcommand{\epsB}{\varepsilon_{\rm B}}
\newcommand{\epsnu}{\varepsilon_{\rm \nu}}
\newcommand{\epsCMB}{\varepsilon_{\rm cmb}}

\newcommand{\ale}{\alpha_{\rm e}}
\newcommand{\as}{a_{\rm s}}
\newcommand{\ab}{a_{\rm b}}
\newcommand{\aC}{a_{\rm C}}

\newcommand{\hq}{\hbar}

\begin{document}

\thesaurus{ 
02.18.5; 	
02.19.2;	
11.03.4: Coma;	
11.09.3;	
11.13.2	
}

\title{Non-thermal Origin of the EUV and HEX Excess Emission
of the Coma Cluster - the Nature of the Energetic Electrons}
\author{Torsten A. En{\ss}lin\inst{1}, Richard Lieu\inst{2}, Peter L.
Biermann\inst{1} } 
\institute{Max-Planck-Institut f\"{u}r Radioastronomie, Auf dem 
H\"{u}gel 69, D-53121 Bonn, Germany \and Department of Physics, University of
Alabama, Huntsville, AL 35899, USA\\ensslin@mpifr-bonn.mpg.de,
lieur@cspar.uah.edu, plbiermann@mpifr-bonn.mpg.de}
\date{Received ??? , Accepted ???}  
\maketitle\markboth{En{\ss}lin, Lieu \& Biermann: Non-thermal Origin
of Excess Emission of the Coma Cluster}{En{\ss}lin, Lieu \& Biermann:
Non-thermal Origin of Excess Emission of the Coma Cluster}

\abstract{The excess in extreme-ultraviolet (EUV) radiation and the recently
discovered high energy X-ray (HEX) excess from the Coma cluster may be
modeled using fewer parameters than in a thermal gas scenario, yet
equally satisfactorily, by power law spectra.  Their origin could
therefore be inverse-Compton (IC) emission by relativistic
electrons. The scattered background photon field can either be the
cosmic microwave background (CMB) or the starlight of the elliptical
galaxies within the cluster.

For the EUV excess both possibilities are consistent with the present
data. If the EUV excess is due to CMB scattering, a strongly
inhomogeneous magnetized intra-cluster medium (ICM) is required, in
which the density of the IC scattering relativistic electrons is
anticorrelated with the magnetic field. This could be understood if
the electrons were accelerated during a major merger event within the
last 2 Gyr and cooled afterwards in the inhomogeneous fields. If the
EUV excess is due to scattered starlight, a population of
relativistic, very low energy electrons has to be present, which would
have a high energy density. In order to survive Coulomb losses, these
electrons have to be separated from the dense thermal cluster gas by
confining magnetic fields. Such a second component of the ICM could be
remnant radio plasma left over from the epoch of violent quasar
activity, which did not mix with the ICM. The observed narrow radial
profile of the EUV excess emission is a natural consequence of this
model due to the narrow profile of the photon distribution.  Both
models favor therefore very inhomogeneous magnetic field and
relativistic electron distributions.

The IC models for the HEX excess require implausible conditions. CMB
scattering leads to a mainly unmagnetized ICM, in contradiction to Faraday
rotation measurements. Starlight IC scattering electrons would overproduce
EUV photons due to simultaneously CMB scattering. We propose that the
observed HEX excess is due to bremsstrahlung of a small high energy
power-law tail of the mainly thermal ICM electron distribution. Such a tail
is expected since some degree of turbulence has to be present within the
ICM, which naturally accelerates electrons out of the thermal pool.

}

\section{Introduction}

The intra-cluster medium (ICM) of clusters of galaxies is known to consist of
hot gas with temperatures of several keV. The gas is visible through the
X-ray bremsstrahlung emitted by the hot electrons.  Recent measurements in
the spectral range below and above the thermal bulge show excess emission
above what is expected from the thermal. This could indicate the presence of
regions with different temperatures within the ICM, which could only be
present there if the thermal conductivity is strongly reduced e.g. by tangled
magnetic fields. But the excess emissions could also trace energetic
non-thermal electrons via non-thermal radiation processes as bremsstrahlung
and inverse Compton (IC) scattering of background photons.

We discuss three different non-thermal processes which could lead to the
extreme ultraviolet (EUV) excess and the high energy X-ray (HEX) excess and
their physical implications for the conditions within the ICM: In
Sec.~\ref{sec:CMB-IC} the possibility that IC scattering of CMB photons is
responsible for the observed excess emissions is examined. In
Sec.~\ref{sec:SL-IC} we describe a model for anisotropic IC scattering, as it
has to be considered if the excess emissions are the scattered anisotropic
starlight within the cluster. And in Sec.~\ref{sec:Brems} non-thermal
bremsstrahlung is proposed as an explanation of the HEX excess. In order to
keep the discussion of the various IC processes as clear as possible, we
concentrate on the EUV excess, and explain the consequences of the HEX excess
for the different models in separate subsections.  Sec.~\ref{sec:disc}
contains a concise discussion of our results.

Parameters are estimated for $H_{\rm o} = 50\,{\rm km\,s^{-1}\,
Mpc}\,h_{50}$, and $q_o = 0.5$, so that at the distance of Coma $1'
\cong 39\, {\rm kpc}\,h_{50}^{-1}$.

\subsection{EUV Excess}

\begin{table}
\begin{center}
\begin{tabular}{ccc}
\hline
region & spectral index & norm \\
arcmin & & $10^{-3} {\rm cm^{-2}\,s^{-1}\,keV^{-1}}$\\
\hline\\[-0.6em]
0--3&       $1.75^{+0.26}_{-0.18}$& $1.17^{+1.07}_{-0.58}$  \\
3--6&       $1.73^{+0.31}_{-0.16}$& $2.12^{+1.74}_{-1.15}$ \\
6--9&       $1.73^{+0.26}_{-0.16}$& $2.72^{+2.10}_{-1.36}$ \\
9-12&       $1.81^{+0.56}_{-0.25}$& $1.65^{+2.25}_{-1.16}$ \\
12--15&     $1.70^{+0.43}_{-0.19}$& $1.62^{+1.98}_{-1.06}$ \\
15--18&     $1.69^{+0.74}_{-0.22}$& $1.58^{+2.65}_{-1.28}$ \\
\end{tabular}
\end{center}
\caption[]{\label{tab:data}Results of the re-modeled EUV and soft X-ray data
by Lieu et al. (1999). For the different angular rings the
differential number index of the photon excess flux distribution, and
the normalization (extrapolated flux at 1 keV) is given.}
\end{table}

The origin of the EUV excess observed in some clusters of galaxies
(\object{Coma}, \object{Virgo}, \object{Abell 1795}, \object{Abell 2199})
is still under discussion. The first reports of this emission interpreted
it in terms of a relatively cool component ($<$ keV) of the hot ICM (Lieu
et al. 1996a, 1996b; Bowyer et al. 1996, 1997; Mittaz et al. 1998). Doubts
about the emission from cold gas were given by the nondetection of
resonance lines by Dixon et al.  (1996), expected in this case. An
alternative explanation by inverse Compton (IC) scattered background light
was proposed (Hwang 1997, En{\ss}lin \& Biermann 1998, Sarazin \& Lieu
1998, Bowyer \& Bergh\"ofer 1998). The reader is referred to Sarazin \&
Lieu (1998) and Bowyer \& Bergh\"ofer (1998) for a broader discussion of
the literature.

\begin{figure}
\psfig{figure=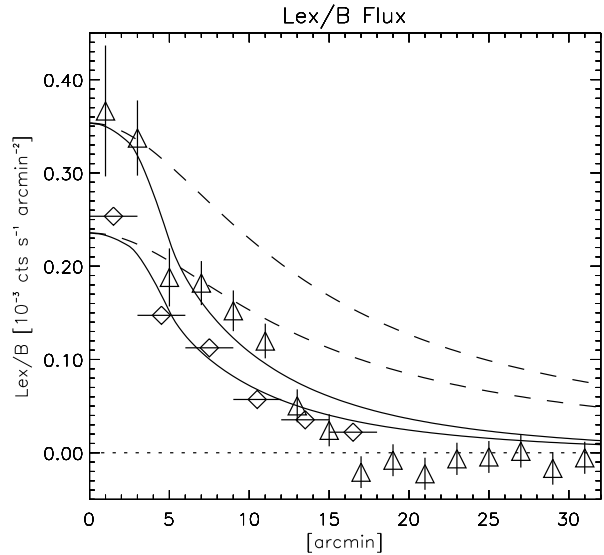,width=0.9\textwidth,angle=0}
\caption[]{\label{fig:LexB} Observed Lex/B excess flux from Bowyer and
Bergh\"ofer (1998) (triangles) and Lieu et al. (1999) (diamonds).
The solid lines are the prediction of the anisotropic starlight-IC
model of Sec.~\ref{sec:SL-IC} for the two normalizations given there.
The dashed lines are scaled projected electron profiles. The
difference between the solid and dashed lines is due to the central
enhancement in starlight photon density. The theoretical curves would
be lower at larger radii, if the effect of enhanced IC scattering near
galaxies exists as discussed in the {Appendix}. Note that the
systematic uncertainties in the observed profile increases with radius
due to the difficult data analysis.}
\end{figure}

Recently, Lieu et al. (1999) re-modeled the soft excess in Coma
with a power-law emission spectrum, and found satisfactory fits
(reduced $\chi^2$ between 1.10 and 1.36 for 180 degrees of freedom).
Tab.~\ref{tab:data} excerpts those aspects of their results relevant
to this paper.  The differential number index of the photon
distribution is nearly independent of radius, so that we use a
constant value of 1.75 for comparison with theoretical models. The
spatial emission profile is narrow, and is plotted in
Fig.~\ref{fig:LexB} in comparison with the profile estimated by Bowyer
and Bergh\"ofer (1998), which is higher by 50\%, and in comparison
with the result of the starlight-IC model described in
Sec.~\ref{sec:SL-IC}.

If IC scattering is the process producing the excess, as will be
assumed in this paper, then relativistic electrons have to be present
within the ICM. This is not surprising, since the Coma cluster is
known to contain one of the largest radio halos (Willson 1970), and has
therefore relativistic electrons in the energy range of $(1.2 - 3.6)$
${\rm GeV}$ $(B/(6 \,\mu{\rm G}))^{-1/2}$, depending on the field
strength $B$. The energy range of the electrons producing the IC
emission can be estimated from the fact that electrons with momentum
$P_{\rm e}$ scatter the peak of a thermal photon population to an
average energy of $<\varepsilon > =
\frac{4}{3}\,p^2\, 2.70\,k_{\rm B}T$ (Blumenthal \& Gould 1970)\nocite{blumenthal70},
where $ p = P_{\rm e}/(\me c)$.  Scattering of the CMB into the observed
range of 69--400 eV requires therefore that the electrons have momentum in
the range 140 -- 350 MeV/c. Another possibility is that the starlight of
elliptical galaxies is scattered into the observational band (En{\ss}lin \&
Biermann 1998). Since the temperature of this radiation field is roughly
3000 K, required electron momenta are in the range of only 4.4 -- 10.6
MeV/c.

If it is possible to decide from physical consideration, which photon
distribution is scattered into the EUV range and is observed, immediately
information is given about a part of the relativistic electron spectrum
below the radio range.  The differential number index of the electron
population in the energy range responsible for the emission has to be $\ale
=2.5$, in order to produce an observed IC flux with photon differential
number index of $1.75$.

\subsection{HEX excess\label{sec:HEXE}}

\begin{figure}
\psfig{figure=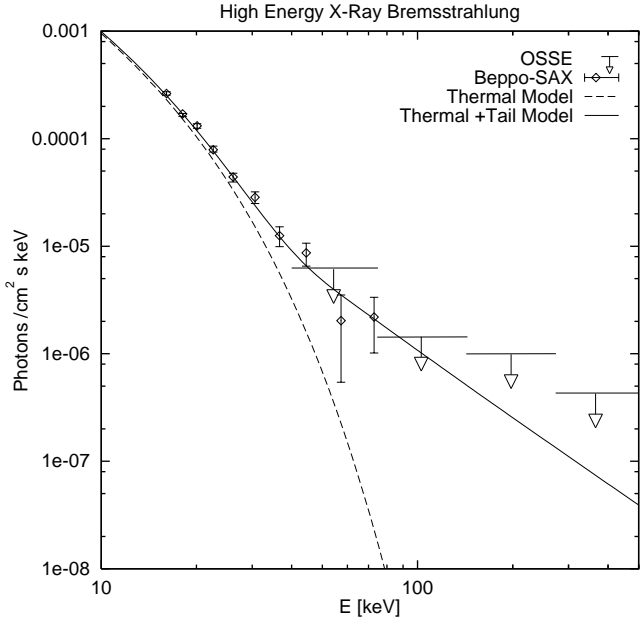,width=0.5\textwidth,angle=0}
\caption[]{\label{fig:HEX} HEX emission of Coma. The data points are
from the Beppo-SAX measurement (Fusco-Femiano et al. 1998, 1999) and the upper
limits from OSSE (Rephaeli et al. 1994). The lines are theoretical
spectra, calculated in Sec.~\ref{sec:Brems}. The dashed curve belongs to a
thermal electron distribution, and the solid curve if a non-thermal tail with
number index $\ale =2.5$ is appended above a momentum of $P_{\rm e} =
0.5\,\me\,c$ (or an kinetic energy of 60 keV).}
\end{figure}

Early attempts to measure non-thermal HEX emission from the Coma
cluster gave only upper limits, or inconsistent results (see Rephaeli
et al. 1994 for upper limits of the OSSE experiment and for further
references). A recent measurement by Fusco-Femiano et al.  (1998,
1999) with the Beppo-SAX satellite detected an excess emission above
25~keV, consistent with the OSSE upper limits. Assuming that the
excess is due to a second thermal component would require a
temperature of 40 keV for it (Fusco-Femiano et al. 1998, 1999),
which seems to be implausible.  The HEX spectrum is plotted in
Fig. \ref{fig:HEX} in comparison with a thermal, and a modified
thermal spectrum.

If the HEX is due to IC scattering of CMB photons, the necessary electrons
would be within the energy range of 2.8--4.9 GeV, and therefore possibly
visible at radio frequencies. In order to scatter starlight photons into the
observed energy band, electrons between 84 and 150 MeV are
needed. Fusco-Femiano et al. (1998) report that the HEX excess can be fitted
satisfactorily with a power-law with photon number index of 0.97--3.45. The
HEX excess flux is $2 \cdot 10^{-11}\, {\rm erg\, cm^{-2}\, s^{-1}}$ between
20 and 80 keV, and within a radius of $1^\circ$. For comparison: An
extrapolation of the EUV spectra (Tab. \ref{tab:data}) to higher energies
gives an excess flux between 20--80 keV of $1\cdot 10^{-11}\, {\rm erg\,
cm^{-2}\, s^{-1}}$ within a radius of 18'. Since the difference by a factor
of two might be due to the larger field of view in the second case, a
variation or a systematic error in the spectral indices, or just a systematic
effect in the estimate of the excess emission (e.g. Bowyer \& Bergh\"ofer
(1998) get a higher EUV excess by 50~\%) the agreement suggests a possible
physical connection.

\section{\label{sec:CMB-IC}CMB-IC}
\subsection{Difficulties with the radio halo electrons\label{sec:diff}} 

Hwang (1997) and En{\ss}lin \& Biermann (1998) discussed an inverse
Compton model for the EUV excess in which the CMB photons are
scattered by a population of relativistic electrons, which are the low
energy tail of the population observed in the radio halo of Coma. This
model is problematic for three reasons:

{\bf First}, the spectral index of the radio emission is 1.16
(excluding some high frequency data point, Bowyer \& Bergh\"ofer
1998). Thus, the synchrotron emitting electrons have a differential
number index of $\ale = 3.32$, steeper than a CMB-IC scattering
component with $\ale =2.5$. This means that a break has to be
present in the electron spectrum at lower than radio emitting
energies if both components belong to the same population.  Giovannini
et al. (1993) report a radial spectral index decrease of the radio
halo. The central spectral index is 0.8 and the outer 1.8. The first
corresponds to an electron number index of 2.6, which would be in good
agreement with the required number index of the EUV producing
electrons. Although this solves the first problem, it increases the
difficulties with the magnetic field estimate as explained below.

{\bf Second}, Bowyer and Bergh\"ofer (1998) pointed out, that the
radial profile of the EUV excess has a full width half maximum (FWHM)
of $15'\!\!.8$ ($19'\!\!.3\times 12'\!\!.6 \pm 1'\!\!.5$). If the EUV
excess is due to scattered CMB photons, the necessary electrons,
which have energies of $140 - 350$ MeV, have a profile with the same
FWHM as the excess emission, whereas the low frequency radio profile
has a much broader profile with a FWHM of $\approx 24'$. The strength
of the radio emission, resulting from electrons with energies of $(1.2
- 3.6)$ ${\rm GeV}$ $(B/(6\,\mu{\rm G}))^{-1/2}$, is a product of the
spatial densities of these electrons and $B^2$ (approximately). Since
any reasonable profile of the magnetic fields should decrease with
radius on the scale of a core radius, the spatial profile of these
radio electrons has to be even broader than the radio emission itself,
which is broader than the EUV emission. Thus, the FWHM of the electron
population has to drop from a value which is considerably larger than
$\approx 24'$ at 1 GeV (the radio range) to $15'\!\!.8$ at 350 MeV
(CMB-IC scattering electrons). In other words, a low energy
electron population which is spatially very differently distributed
compared to the population producing the radio halo is required
(Bowyer \& Bergh\"ofer 1998).

An extrapolating of the radial dependent radio spectra of Giovannini et
al. (1993), would give an electron population at lower energies, which
is less centrally concentrated as the radio population, due to the
flat central spectral index. This is even more in conflict with the
required compact spatial distribution of the EUV electrons.

{\bf Third}, matching the distribution of the CMB scattering electrons to
that of the radio emitting ones, in the sense that both populations belong
to the same power law, requires that the central magnetic field strength in
the ICM is $\approx 1 \mu$G, otherwise the total number of relativistic
electrons would be too small in order to produce the EUV excess flux
(En{\ss}lin \& Biermann 1998, Bowyer \& Bergh\"ofer 1998, see also Hwang
1997, who derives a volume averaged field strength of 0.4 $\mu$G). Since a
spectral break is necessary between the regions of different spectral
indices of radio- and IC-electron population, the allowed magnetic field
strength is even lower than 1 $\mu$G.

The magnetic fields of the ICM can be independently measured by Faraday
rotation of linear polarized radio emission traversing the ICM. The main
uncertainty of this method is the underlying magnetic field reversal scale. A
scale of 10 kpc$\,h_{50}^{-1}$ gives a central field strength of $1.7\pm
0.9\, \mu$G$\,h_{50}^{1/2}$ (Kim et al. 1990). But high resolution
depolarization measurements indicate that the reversal scale is 1
kpc$\,h_{50}^{-1}$ or below, leading to a ICM field strength of at least
$6\pm 1\, \mu$G$\,h_{50}^{1/2}$ (Feretti et al. 1995, but also predicted by
Crusius-W{\"a}tzel et al. 1990). The field strength seems therefore to be
higher than allowed by this IC model. The discrepancy between the field
strength from Faraday rotation measurements and that from the synchrotron/IC
ratio is solved,
\begin{enumerate}
\item[(a)]
if there is a sharp step in the electron spectrum {between} the IC
emitting electrons (140--350 MeV), and the radio emitting electrons
(above 1 GeV), or
\item[(b)]
if the medium is inhomogeneously magnetized on a small scale compared to the
observational spatial resolution, and there is a strong anticorrelation
between magnetic field strength and density of relativistic electrons -- as
En{\ss}lin \& Biermann (1998) proposed.
\end{enumerate}
We demonstrate in the following, that for an electron population
cooling in an inhomogeneously magnetized medium both are the case.

\subsection{Merger events as a source of relativistic electrons}

Such a population of electrons could initially be produced by shock
acceleration during an energetic merger event in the past of Coma. Several
such events should have happened, and there is evidence for recent and
on-going events: Burns et al.~(1995)\nocite{burns95} propose that the X-ray
emitting blob south-west of the center of Coma is due to the ascending motion
of a group of galaxies around NGC 4839, which had a core passage 2 Gyr
ago. Using kinematical considerations Colless \& Dunn (1996) argue that this
group is most probably infalling, therefore being in a pre- instead of
post-merger stadium. Nevertheless these authors and others (Biviano et
al. 1996, Vikhlinin et al. 1997, Donnelly et al. 1998) find apparent evidence
for on-going or recent merger events in the cluster core from the galaxy
velocity distribution and X-ray substructure. The time of injection of a
relativistic electron population is expected to be during the presence of a
strong shock wave in the ICM. Therefore an acceleration of a relativistic
electron population within the last 2 Gyr is reasonable.

Synchrotron and inverse Compton cooling would lead to a cutoff in this
distribution which evolves with time from higher to lower energies and
would have reached 300 MeV after less than 1 Gyr for a magnetic field
strength of 6$\,\mu$G. After 1.5 Gyr it would be below 150 MeV (see
Fig.~\ref{fig:epop}). The injection of electrons which currently
scatter the CMB into the EUV range should have happened less than 1
Gyr ago if the magnetic fields are homogeneous at $6\,\mu$G. But for
inhomogeneous fields, and for an injection age of up to 2 Gyr
sufficient electrons are still in this energy range, as will be
demonstrated in the next section.

\subsection{Cooling electrons in an inhomogeneous medium\label{sec:cool}}

\begin{figure*}
\psfig{figure=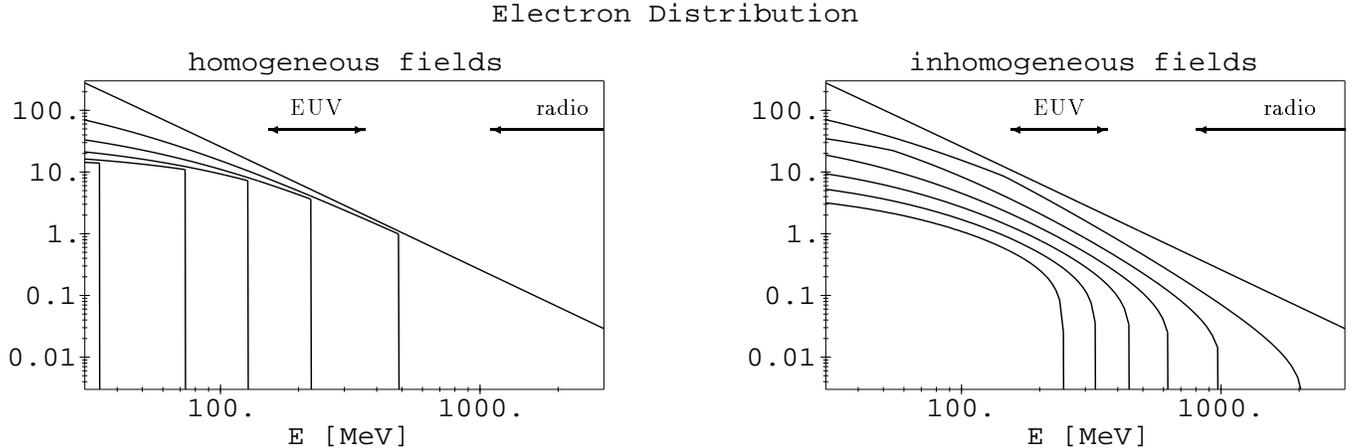,width=\textwidth,angle=0}
\caption[]{\label{fig:epop}Cooling electrons in a homogeneous magnetic
 field of 6 $\mu$G (left) and in inhomogeneous fields with mean field
strength of 6 $\mu$G (right) ($\tilde{f}_{\rm e}(p,t)\, dp$). In the latter
case the average electron population is shown, averaged over the populations
at different field strengths. The initial electron spectrum has a
differential number index of 2 and is the highest curve. Plotted below this
are electron populations after 0.5, 1, 1.5, 2, 2.5, and 3 Gyr of cooling. The
energy ranges mainly responsible for the EUV excess emission by scattering of
CMB photons and for radio emission are indicated. The different radio ranges
result from different magnetic field strength distributions in these
models. In the inhomogeneous model, the radio range has to be used with care,
since the shown electron distribution is integrated over all regions with
different field strength, but the radio range is shown for a field strength
of $12\,\mu$G.}
\end{figure*}

A relativistic electron with dimensionless momentum $p = P_{\rm e}/(\me\,c)$
loses energy/momentum (Kardashev 1962) in the ICM at a rate of
\begin{equation} 
\label{eq:cool}
- \frac{dp}{dt} =  \aC + \ab p + \as p^2.
\end{equation}
The coefficient $\aC$, $\ab$, and $\as$ for Coulomb, bremsstrahlung, and
synchrotron/IC losses are (Rephaeli 1979\nocite{rephaeli79}, Blumenthal
\& Gould 1970\nocite{blumenthal70})
\begin{eqnarray}
\aC &=& \frac{3}{2} \sigma_{\rm T} \,c\, n_{\rm e} \left( \ln \frac{\me
c^2 p^{1/2}}{ \hq \omega_{\rm p}} +0.22\right),\\
\ab &=& \frac{3\alpha}{\pi} \sigma_{\rm T} \,c\, n_{\rm e} \left(\ln
2p -\frac{1}{3} \right),\\
\label{eq:as}
\as &=& \frac{4}{3} \sigma_{\rm T} \,c\, \frac{\epsB + \epsCMB}{\me c^2},
\end{eqnarray}
where the electron density of the background gas is $n_{\rm e}$, $\epsB$ is
the magnetic field and $\epsCMB$ the CMB photon energy density.  The plasma
frequency is $\omega_{\rm p} = \sqrt{4 \pi e^2 n_{\rm e}/\me}$ and $\alpha$
is the fine-structure constant. $\aC$ and $\ab$ depend weakly on $p$, which
we neglect in the following by inserting a typical value of $p = 10^3$ into
the logarithms. We also use a density of $n_{\rm e} = 3\cdot 10^{-3} \,{\rm
cm^{-3}}\,h_{50}^{1/2}$ (Briel et al. 1992\nocite{briel92}) for the plasma
frequency and in our following examples. Only a small error is introduced by
using the constant $p$ within the logarithms, but it allows an analytical
calculation of the time dependent electron distribution. Further we assumed
sufficiently fast pitch angle scattering of the electrons, so that the
distribution can always be assumed to be isotropic, and Eq.~\ref{eq:as}
holds.  The time needed to cool from $p_0$ to $p_1$ is given by an
integration of Eq.
\ref{eq:cool}
\begin{eqnarray}
t_{\rm cool}(p_0,p_1) = \frac{1}{\as p_*}
&& \left[\arctan(\frac{\ab}{2\as p_*}+
\frac{p_0}{p_*}) - \right. \nonumber\\
&&\left. \arctan(\frac{\ab}{2\as p_*}+
\frac{p_1}{p_*})\right], 
\end{eqnarray}
with 
\begin{equation}
p_* = \frac{\sqrt{4 \as \aC - \ab^2}}{2\as},
\end{equation}
so that any electron loses all its energy in a time shorter than
\begin{equation}
t_{\rm cool}(\infty,0) =
\frac{1}{\as  p_*} \left( \frac{\pi}{2}- \arctan
( \frac{\ab}{2\as p_*}) \right) .
\end{equation}
This is 3 Gyr for a field strength of 6 $\mu$G and a gas electron density
of $n_{\rm e} = 3\cdot 10^{-3} \,{\rm cm^{-3}}$.
An electron with $ p_1$ at the time $t$, had the initial momentum
\begin{equation}
 p_0( p_1,t) =  p_* \tan \left( \as  p_* t +
\arctan ( \frac{\ab}{2 \as  p_*}  +\frac{ p_1}{ p_*} )
\right)-\frac{\ab}{2 \as}.
\end{equation}
Thus, a population of relativistic electrons $f_{\rm
e,0}( p_0)\,d p_0$ which was injected at $t=0$ into the ICM
and which cooled afterwards without any additional acceleration has a
time dependent distribution function, which is simply given by a
transformation of the initial
distribution under the mapping $p_0 \rightarrow p_1$
\begin{equation}
\label{eq:fe1}
f_{\rm e}( p_1,t) = f_{\rm e,0}( p_0( p_1,t))\,
\frac{d p_0( p_1,t)}{d p_1}
\end{equation}
as long as $t\leq t_{\rm cool} (\infty, p_1)$, otherwise $f_{\rm
e}( p_1,t) =0$.

In an inhomogeneously magnetized medium the spectrum of the cooling
electrons becomes a function of position even in the case the initial
spectrum was homogeneous due to the spatial dependence of the
cooling. We suppose that there is no significant exchange of electrons
between the different magnetized regions. This is a necessary
condition in order to establish the required anticorrelation of
magnetic fields and the electron population. It is also reasonable if
the origin of the magnetic fields was injection by radio galaxies
(Daly \& Loeb 1990; En{\ss}lin et al. 1997, 1998) or by galactic winds
(Kronberg et al. 1999, but see also V\"olk et al. 1996 and
V\"olk \& Atoyan (1998) for the amount of non-thermal energy injected
by galactic winds), since then the topology of the fields can be
expected to be closed. Electrons have to diffuse perpendicular to the
field lines in order to reach a region with different field strength,
which is a very slow process. In case of in-situ generated fields an
inhomogeneous medium is also possible as the sun probably
demonstrates.

\begin{figure*}
\psfig{figure=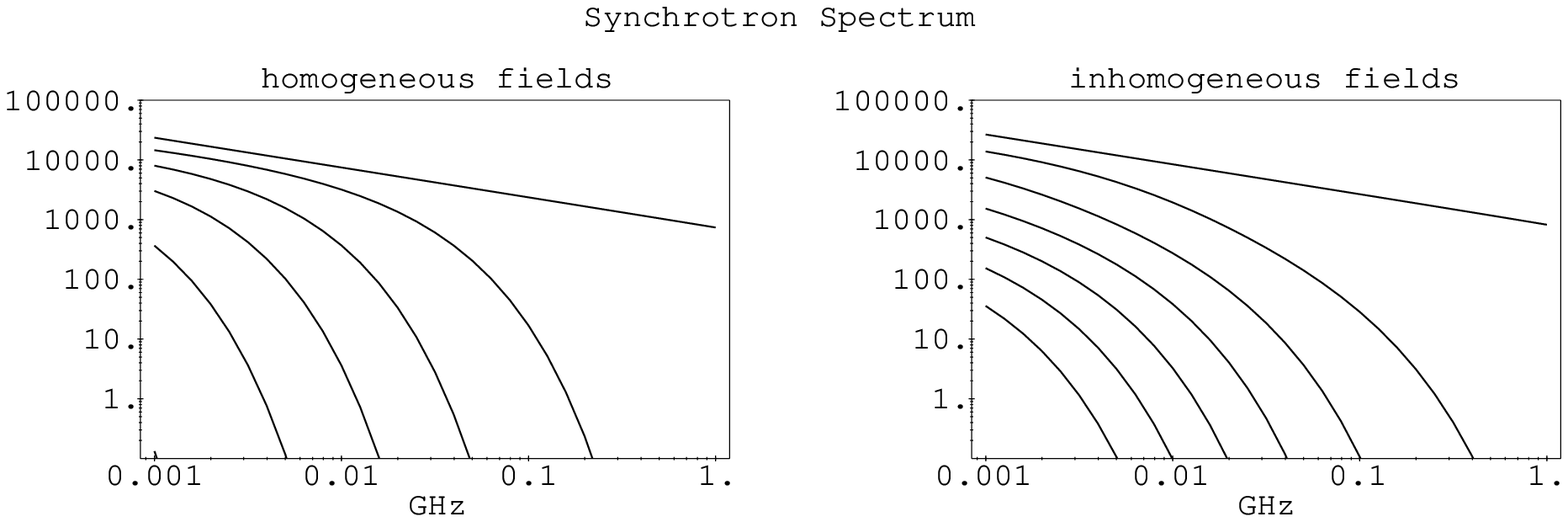,width=\textwidth,angle=0}
\caption[]{\label{fig:sync} Synchrotron emission of electron
populations shown in Fig.~\ref{fig:epop} at an age of 0, 0.5,
1, 1.5, 2, 2.5, and 3 Gyr in units of $c_3 \,\mu$G$ = 1.87 \cdot
10^{-29}$ erg s$^{-1}$ Hz$^{-1}$.}
\end{figure*}

Assuming a distribution of field strength $f_B (B) \, dB$ within a
volume, which is sufficiently large compared to typical sizes of
the magnetized regions, but small enough to be spatially unresolved
by the current observations, we get a volume averaged electron
population
\begin{equation}
\label{eq:feav}
\tilde{f}_{\rm e}( p,t) = \int dB\, f_e( p,t,B) \,\,
\end{equation}
where we wrote $f_{\rm e}( p,t,B)$ instead of $f_{\rm
e}( p,t)$ in order to show explicitly the dependence of the local
electron population on the local field strength. This averaged population is
directly seen by the IC flux. Fig.~\ref{fig:epop} shows the volume
averaged distribution function $\tilde{f}_{\rm e}( p,t)$ for an
injected distribution
\begin{equation}
f_{\rm e, 0}( p) = C_{\rm e}\, p^{-2}, \mbox{ with } C_{\rm
e}= 10^6, 
\end{equation}
as one would expect after particle acceleration due to a strong shock wave in
the ICM of a cluster merger event. We assume a normalized distribution of
magnetic fields
\begin{equation}
f_B(B) = 1/B_{\rm max};\;\;\mbox{for }\;\;0\mu{\rm G} < B < B_{\rm max}=
12\mu{\rm G},
\end{equation}
and $f_B(B)=0$ otherwise. This choice is arbitrary, but demonstrates well the
effects we want to discuss. It leads to an averaged field strength of
$6\,\mu$G consistent with Faraday rotation measurements. The electron
population in a homogeneous field
\begin{equation}
f_B(B)= \delta(B-6\mu{\rm G})
\end{equation}
is also shown in Fig.~\ref{fig:epop} for comparison.

The synchrotron emissivity at a given frequency $\nu$ of an isotropic
distribution of electrons in a
randomly oriented distribution of magnetic fields is
\begin{equation}
\epsnu (t) = c_3 \int dB f_B(B)  B\int d p f_{\rm e} ( p,t,B)
\tilde{F}(\nu/ \nu_{\rm c}),
\end{equation}
where $c_3 = \sqrt{3} e^3/(4 \pi \me c^2) $, $\nu_{\rm c}( p,B)=3 e B
p^2 /(4\pi \me c)$, and the angle averaged dimensionless spectral
emissivity of a monoenergetic isotropic electron distribution (Crusius
\& Schlickeiser 1986) is
\begin{equation}
\tilde{F}(x) = \frac{\pi\,x }{2} \left(W_{0,\frac{4}{3}}(x)
\,W_{0,\frac{1}{3}}(x) - W_{\frac{1}{2},\frac{5}{6}}(x)\,
W_{-\frac{1}{2},\frac{5}{6}}(x) \right) ,
\end{equation}
with $W_{\rm \lambda,\mu}(x)$ denoting Whittaker's function (Abramowitz \&
Stegun 1965\nocite{abramowitz65}). $\tilde{F}(x)$ can be approximated to an
accuracy of a few percent by
\begin{equation}
\tilde{F}(x) \approx \frac{2^{2/3} }{ \Gamma(11/6) }
\left(\frac{\pi}{3}\right)^{3/2} x^{1/3} \exp \left( -\frac{11}{8}
x^{7/8}\right), 
\end{equation}
which is faster to evaluate numerically.  Fig.~\ref{fig:sync} shows the
synchrotron emission of a cooling population of electrons in the two models
given above.

The important result of this simple inhomogeneous model is that the electron
population in the weak field regions is still present in the energy range of
140 -- 350 MeV after 2 Gyr of cooling, so that it can produce the EUV excess
by IC scattering of CMB photons. But synchrotron/IC cooling produces a sharp
cutoff at higher energies, so that for any given normalization of the
electron population no observable radio emission remains. In the case of
homogeneous fields a considerable amount of fine tuning of the time point of
injection is necessary in order to allow a large electron population in the
140 -- 350 MeV range, without overproducing low frequency radio emission,
and therefore violating the observational constraints. As long as the
magnetic fields are not lower than $6\,\mu$G, the cooling time has to be near
0.6 Gyr, otherwise either the cutoff is lower than 400 MeV, or the electrons
are visible in the radio.

An electron differential number index of 2.5 for the EUV producing energy
range can easily be matched for sufficiently flat injection spectra.
Unfortunately, the present day differential number index depends not only on
the injection index and the cooling time, but also sensitively on the
distribution of field strength. For the uniform distribution of field
strength assumed above an injection index of $2.15$ would lead to a present
day index of $\approx 2.5$ in this energy range. If the field distribution is
more strongly weighted to lower field strength, a steeper injection spectral
index would be required, and if fewer low magnetic field regions exist a
flatter one. With our poor present day knowledge about the distribution of
field strengths it is therefore impossible to derive the injection spectrum.

The energy density of the relativistic electrons in the inhomogeneous model
above after 2 Gyr is nearly two orders of magnitude smaller than the
injected energy density, assuming that the injected spectrum extends from
$P_{\rm e} = 1$ MeV/c to $10^4$ MeV/c, and the injection differential number
index is 2.15. This energy loss depends strongly on the assumed distribution
of fields: if more weak field regions are present the energy loss is much
less dramatic. And since the present day population of EUV excess producing
electrons has an energy density which is a factor of a few hundred below the
thermal energy density (see Fig. 5. in Bowyer \& Bergh\"ofer 1998), an
injection energy density considerably lower than the present thermal energy
density would be sufficient to explain the present day EUV excess.

The effects of radius dependent electron and magnetic field
distributions can in principle be treated with a similar formalism,
where the volume average is then over elongated beams along the line
of sight. Instead of giving a detailed model, which would rely on even
more assumptions, we briefly discuss the qualitative behaviour. The
magnetic fields strength and the injected electron energy density
should decrease with radius.  Due to the higher fraction of strong
field regions in the center, one would expect a stronger cooling of
the electrons there, and therefore a lack of energetic electrons
there after a typical cooling time. But the fate of the low energy
electrons, necessary for the CMB-IC EUV production, is mainly
determined by the volume fraction of the low field strength
regions. If e.g. their volume fraction is constant with cluster radius
(but the field strength of the strongly magnetized region is still
varying with radius), then these electrons are still numerous in
the cluster center.

\subsection{In-situ accelerated electrons\label{sec:in-situ}}

Another possible origin of an electron population seen by CMB-IC is
in-situ acceleration by plasma waves. A strong wave field can be
expected if there is an on-going merger event in the center of Coma,
as several authors (Colless \& Dunn 1996, Biviano et al. 1996,
Vikhlinin et al. 1997, Donnelly et al. 1998) report. The electron
spectrum produced can be approximated by
\begin{equation}
f_{\rm e}( p)d p = C_{\rm e} \, p^{-\ale}\, \exp(
-\frac{ p}{ p_{\rm c}}) \,dp
\end{equation}
(Schlickeiser 1984\nocite{schlickeiser84}),
where the cutoff 
\begin{equation}
 p_{\rm c} = \frac{v_{\rm A}^2}{9 \as \kappa} \sim
\frac{B^{2+\delta_\kappa}}{B^2 + 8 \pi \epsCMB}
\end{equation}
depends on the ratio of acceleration time scale to synchrotron/IC
cooling time. $v_{\rm A}$ is the Alfv\'en velocity and $\kappa \sim
B^{-\delta_\kappa}$ the spatial diffusion coefficient
($0\leq\delta_\kappa\leq 1$). The cutoff $p_{\rm c}$ is a
monotonically increasing function of the magnetic field strength,
which implies that in the case of inhomogeneous fields the in-situ
accelerated electron population reaches highest energies in the
regions of strongest field strength.  This is also supported by the
more complicated dependence of $\ale$ on the magnetic field strength:
stronger fields result in harder spectra. The property of spatial
anticorrelation between electron and magnetic field distributions,
allowing a high number of radio quiet, low energy electrons, can
therefore not be achieved within an in-situ acceleration model. In the
case of inhomogeneous fields the in-situ acceleration model is more
constrained than in the homogeneous case, due to the correlation of
in-situ accelerated electrons and magnetic fields. But even for
homogeneous fields the optimal parameters $\ale=1$, $ p_{\rm c} =
280$, which lead to an electron differential number index of 2.5
between 150 and 300 MeV/c, and which are still allowed by the theory
of in-situ acceleration developed by Schlickeiser
(1984)\nocite{schlickeiser84}, lead to an overproduction of
synchrotron emission above 100 MHz for a field strength of 6 $\mu$G
compared to the observations. The reason for this is the relative
softness of the exponential cutoff of the electron population and the
extended spectral width of the synchrotron emission.

We conclude that on-going in-situ acceleration is very unlikely the origin of
a CMB-IC scattering electron population producing the EUV excess. This is
astonishing in the light of evidence for merging activity in the cluster's
center. It indicates that if merger events are responsible for the
acceleration of energetic electrons, this happens only during the early phase
of the merger, probably during the first shock-crossing time.

\subsection{Remaining difficulties of CMB-IC-EUV}

The difficulty of a CMB-IC model for the EUV excess of Coma with the
discrepancy between magnetic field strength observed by Faraday rotation,
and the estimate using the IC- and synchrotron emission, can be overcome in
a model where electrons cool in an inhomogeneous magnetic field. The narrow
spatial emission profile is more difficult to understand within such a
model, since one would expect a more extended electron distribution,
in particular since the central electrons should cool faster than the
peripheral ones. In order to decide, if this model is realistic or not,
detailed simulations of the spatial distribution of electron acceleration
during a merger core passage are required.

\subsection{Difficulties of CMB-IC-HEX}

Since the EUV and HEX excess fit roughly into a single power-law the
HEX excess might also be produced by CMB IC scattering. The necessary
electrons should be close to or within the energy range visible in the
radio and therefore both populations have to be at least
similar. Their cooling time due to synchrotron/IC losses is of the
order of $10^8$ years, so that continuous acceleration or very recent
injection into this energy range is necessary.  The observed HEX
emission determines the number of electrons in the radio energy range,
if one assumes the radio index also to be valid for the HEX producing
electrons.  In order to be in agreement with the observed synchrotron
emission, the volume averaged magnetic fields strength has to be 0.16
$\mu$G (Fusco-Femiano et al. 1998, 1999), comparable to 0.4 $\mu$G
given by Hwang (1997) for the EUV emission, which is energetically
more distant to the radio range. The central magnetic field strength
is roughly a factor of 3 higher, leading to $\approx 0.5\,\mu$G, which
seems to be too low in order to be consistent with the Faraday
measurements of $6\,\mu$G$\, \h^{1/2}$.

Also for this 3--5 GeV electron population one might ask if (a) a sharp step
in the electron spectrum, and (b) an inhomogeneous magnetized medium might
resolve this discrepancy.  The ICM needs to consist mainly out of regions
with very weak fields (only a few $0.1\,\mu$G) containing the CMB IC
scattering electrons, which are invisible in the radio due to the weak
fields. But highly magnetized regions have to exist (maybe $10\mu$G or more),
in order to explain the Faraday rotation, which need to contain only few 3--5
GeV electrons. If the difference in electron content of these regions was
established by different cooling the time of injection had to be a few 0.1
Gyr in order to allow the several GeV electrons in the weak field regions
still to be present, but those electrons living in the high field regions to
have cooled to energies invisible in the radio. Both required conditions --
(1) weakly and strongly magnetized regions, but no intermediate $\mu$G
regions, and (2) the recently and necessarily fine tuned injection time --
let this model appear unsatisfactory. We conclude, that IC scattering of CMB
photons as an explanation of the HEX excess seems to be implausible, unless
there is something fundamental wrong with the Faraday rotation magnetic field
estimates.

\section{Starlight-IC\label{sec:SL-IC}}

We propose a second model for the origin of the EUV photons, which has
neither of the difficulties mentioned in Sec.~\ref{sec:diff}: Inverse Compton
scattering of starlight photons by very low energy relativistic
electrons. This possibility was first mentioned by En{\ss}lin and Biermann
(1998)\nocite{ensslin98a}, and used as a restriction on Coma's electron
content in the energy range of a few MeV. The first, and third difficulties
mentioned in Sec.~\ref{sec:diff} do not arise, since these electrons are
energetically far away from the radio emitting range. The second one is also
solved within this model, since in this case not only the electron population
is centrally peaked, but also the target photon distribution, so that the
emission profile can be narrower than the projected electron profile. For
CMB-IC both profiles are the same since the CMB is uniform.

The cooling of electrons with a few MeV within the ICM is dominated
by Coulomb losses and is so strong (cooling times $\leq$ 0.1 Gyr),
\begin{enumerate}
\item[(a)]
that it has to be compensated by efficient in-situ acceleration, or
\item[(b)]
that Coulomb losses have to be suppressed due to confining magnetic fields
which contain the relativistic electrons, but not the dense thermal gas.
\end{enumerate}
The latter might look a little bit artificial on a first view, but since a
huge amount of relativistic plasma was injected into the ICM by radio galaxies
in earlier epochs (En{\ss}lin et al. 1997, 1998), a substantial part of this
plasma might still be there as an unmixed, separate, nearly invisible,
and non-thermal component of the ICM.

\subsection{Anisotropic inverse Compton scattering}

For a given spherical symmetric, radial source function $q_{\rm ph}(r)$
of light emitted by galaxies, the photon density per volume and solid
angle as a function of radius $r$ and $\mu =\cos\theta$, the cosine of
the angle between radial direction and photon direction, is given by a
line integral backwards, over the line where these photons were
emitted:
\begin{equation}
n_{\rm ph}(r,\mu) = \frac{1}{4 \pi c} \int_0^{s_{\rm max}} \!\!\! ds \,
q_{\rm ph}(\sqrt{r^2+s^2 -2 r s \mu }) 
\end{equation}
The maximal distance
\begin{equation}
s_{\rm max} = \mu r +\sqrt{R_{\rm cl}^2 -r^2(1-\mu^2)}
\end{equation}
is determined by the boundary of the cluster, which we assume to be $R_{\rm
cl} = 5$ Mpc$\,h_{50}^{-1}$. The radial emission profile
is that of the galaxy distribution $q_{\rm ph}(r) \sim (1+(r/r_{\rm G})^2)^{-
\alpha_{\rm G}}$, with $\alpha_{\rm G}=0.8$ and $r_{\rm G} = 160 \,{\rm
kpc}\, h_{50}^{-1}$ (Girardi et al. 1995), which we use up to a radius of
$R_{\rm cl}$. The normalization can be obtained by comparison with the
observed central luminosity of elliptical galaxies in Coma, which we
estimate by an integration of the R-band luminosity function of Secker \&
Harris (1996)\nocite{secker96}.  We assume that this radiation has a
blackbody spectrum with a typical temperature of 3000 K, and therefore use a
bolometric correction of $m_{\rm R} - m_{\rm bol} = 1.3$ (Webbink \& Jeffers
1969). We get a central luminosity of $9.3\cdot 10^{12}\, {\rm L_\odot \,
Mpc^{-3}}\,h_{50}$. Not included in this is the contribution of the two
central giant elliptical galaxies NGC 4874 and NGC 4889. Their emission
profile is of course not spherically symmetric with respect to the cluster
center. Nevertheless we approximate their emission of $6.2\cdot 10^{11}\,
{\rm L_\odot}\,h_{50}^{-2}$ (Strom \& Strom 1978)\nocite{strom78} to be
homogeneously smeared out within a spherical shell between the radii of 100
and 200 kpc$\,h_{50}^{-1}$, since the present day EUV observations are not
spatially resolved enough in order to justify the much higher computational
effort a 3-dimensional model would require.

The distribution function of relativistic, low energy electrons is
assumed to have spherical symmetry centered on the optical center
of the cluster:
\begin{equation}
\label{eq:Ceo}
f_{\rm e} (p,r)\,dp = C_{\rm e}(r)\, g(p)\,dp = \frac{C_{\rm e,o}\,
p^{-\ale}\,dp}{ (1+(r/r_{\rm core})^2)^{\frac{3}{2}\beta}} ,
\end{equation}
with $C_{e,o} = 10^{-3} {\rm cm^{-3}}\,h_{50}$, which gives the right
amount of IC photons estimated by Lieu et al. (1999), or $C_{e,o}
= 1.5\cdot 10^{-3} {\rm cm^{-3}}\,h_{50}$ in order to reproduce the
EUV profile of Bowyer \& Bergh\"ofer (1998), which is higher by 50\%
but has a similar slope. Further we use $\ale = 2.5$, $r_{\rm core} =
400$ kpc, and $\beta = 0.75$, which are also the core radius and
$\beta$-parameter of the background gas density (Briel et
al. 1992\nocite{briel92}). Such a profile\footnote{A gaussian profile
$\sim\exp(-r^2/(700 {\rm kpc}\,h_{50}^{-1})^2)$, which is wider than
the observed low frequency radio halo, and would be therefore a
possible profile for the radio electrons, gives reasonable
starlight-IC profiles, especially if one considers the growing
systematic observational uncertainties with radius, and the possible
central enhancement discussed in the {Appendix}.}  would be too narrow
for the radio emitting electrons, but too wide for CMB-IC scattering
electrons. But the profile of the very low energy starlight-IC
electrons, which -- together with the enclosing magnetic fields --
would form a second component of the ICM, might be more similar to the
global ICM gas profile than to that of the high energy radio
electrons. On a small scale, we assume the gas and this non-thermal
component to be separated.

\begin{figure}
\psfig{figure=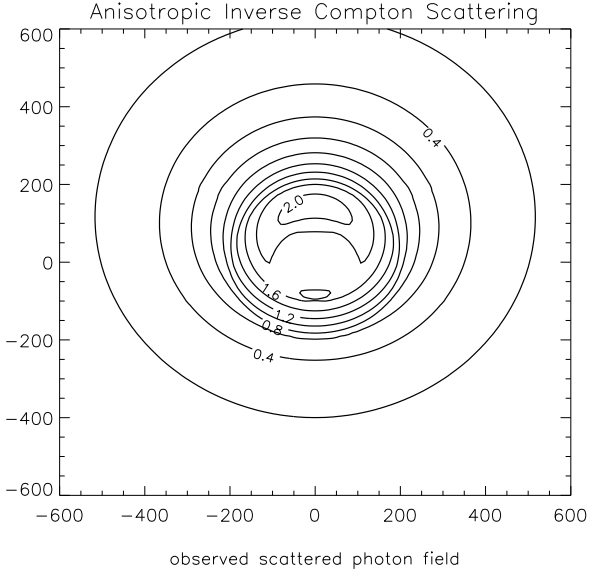,width=0.9\textwidth,angle=0}
\caption[]{\label{fig:slice}Slice through the center of the Coma
cluster. The emissivity into the direction of the observer, who is
located downwards, is shown in units of $10^{-21}$ photons ${\rm
cm^{-3} \, s^{-1}}\,h_{50}$ within the energy band 69--400 eV, and
with the normalization needed to fit the data of Lieu et
al. (1999). The coordinates are given in kpc$\,h_{50}^{-1}$. The
emission profile is not spherically symmetric due to the anisotropic
efficiency of IC scattering by a steep electrum spectrum with
different scattering angle.}
\end{figure}

The production density of IC scattered photons with energy higher
than $E_{\rm ph}$ is
\begin{eqnarray}
q_{\rm ICph}(r,\mu,>E_{\rm ph}) &=& \int_0^{2\pi}\!\!\! d\phi'\,
\int_{-1}^1\!\!\!  d\mu' \,C_{\rm e}(r)\, n_{\rm ph}(r,
\mu') \times \nonumber\\
&&\sigma(\mu',\phi',\mu) \,c\, F(>E_{\rm ph}),
\end{eqnarray}
where $\phi'$ is the azimuth angle between the direction of the
incident photon and the radial direction. The dependence on the
azimuth angle $\phi$ of the scattered photon can be neglected due to
symmetry. The anisotropic cross section of IC scattering by an
isotropic power law distribution of electrons was derived by Brunetti
et al. (1997)\nocite{brunetti97}. The anisotropy of the scattered
photon field is mainly due to the fact, that for fixed initial and
final (observed) photon energies the necessary electron energy depends
on the scattering angle. Thus, for small-angle scattering a higher
electron energy is necessary, which reduces the total number of
scattering events due to the steepness of the electron spectrum.  We
use a crude approximation of the complicated expression given
in Brunetti et al. (1997), which is sufficiently accurate for our
purpose, especially for an electron number index of 3. This is close
enough to the expected 2.5 so that the average deviation is less than
10 \%:
\begin{equation}
\sigma(\mu',\phi',\mu) = \frac{3}{4} \, (\tilde{\mu}(\mu',\phi',\mu)
-1)^2 \sigma_{\rm T} .
\end{equation}
The cosine of the angle between incident and scattered photon is 
\begin{equation}
\tilde{\mu}(\mu',\phi',\mu) = \cos\phi' \sqrt{1-\mu'^2} \sqrt{1-\mu^2}
+ \mu' \mu.
\end{equation}
The spectral slope is given by the standard formula
\begin{eqnarray}
F(>E_{\rm ph}) &\!=\!& \frac{f_{\rm IC}(\ale)}{n_{\rm bb}(k_{\rm B}T)} \,\left(\frac{k_{\rm B}T}{h c}\right)^3
 \left(\frac{E_{\rm ph}}{k_{\rm B}T}\right)^{-\frac{\ale-1}{2}}
\\ 
\nonumber
f_{\rm IC}(\ale) &\!=\!&
\frac{{3 \pi }
2^{\ale+4}(\ale^2+4\ale+11)}{(\ale+3)^2(\ale+5)(\ale^2-1)}
\,\Gamma{\textstyle\left(\frac{\ale+5}{2}\right)}
\,\zeta{\textstyle\left(\frac{\ale+5}{2}\right)}
\end{eqnarray}
(Blumenthal
\& Gould 1970\nocite{blumenthal70}), where $\zeta$ denotes
Riemann's Zeta function, and $n_{\rm bb}(k_{\rm B}T)$ is the thermal
number density of photons in a blackbody cavity with temperature
$k_{\rm B}T$, introduced here for the proper normalization of the
photon production rate. Fig.~\ref{fig:slice} shows a slice through the
cluster center of the emission in the direction of the observer.

The observed scattered photon profile is the line of sight
integration over the scattered photons
\begin{equation}
F_{\rm obs} (R, >E_{\rm ph}) = \int \!dz\, q_{\rm
ICph}(r(R,z),\mu(R,z),>E_{\rm ph}),
\end{equation}
with $r(R,z) = \sqrt{R^2+z^2}$, and $\mu(R,z) =-z/r(R,z)$. The radius
$R$ in this formula can be translated via $1' \cong 39$ ${\rm kpc}$ 
$\h^{-1}$. We note that this anisotropic, line of sight integrated
flux profile differs to one calculated assuming IC scattering to be
isotropic. But our calculation shows that this difference is small.

Correcting for the spectral sensitivity of the Lexan-Boron filter of
the Deep Survey Telescope of the EUV Explorer and the effect of
Galactic absorption for $N_{\rm H} = 8.7\cdot 10^{19} {\rm cm}^{-2}$
(Lieu et al. 1996a), gives the theoretical predicted count rate, which
is shown in Fig.~\ref{fig:LexB} in comparison to the observed profiles
and the slope of the assumed electron profile.  

The IR-flux, produced by these electrons scattering the CMB photons, would be
distributed as broad as the dashed lines in Fig.~\ref{fig:LexB}, and and
would be far below the present day limit on diffuse IR emission from Coma, as
can be seen in Fig. 1 or Tab. 1 in En{\ss}lin \& Biermann (1998).

\subsection{Energy density}

The kinetic energy density of the assumed relativistic electron population
($C_{\rm e,o} = 10^{-3}\,{\rm cm^{-3}}\,h_{50}$, Eq.~\ref{eq:Ceo}) in the
range of $P_{\rm e} = 4.4-10.6$ MeV/c is $1.8\cdot 10^{-10}\, {\rm erg\,
cm^{-3}}\,h_{50}$ at the center of the cluster. This is more than the central
thermal gas energy density of $1.2\cdot 10^{-10}\, {\rm erg\,
cm^{-3}}\,h_{50}^{1/2}$ (Briel et al. 1992\nocite{briel92}). The ratio of the
relativistic to thermal pressure is $0.8\,h_{50}^{1/2}$ due to the smaller
adiabatic index of the relativistic population. This seems to be very high,
since one expects a similar pressure in confining magnetic fields and maybe
in relativistic protons, which would be located at energies around 1 GeV due
to their higher rest mass. Also the lower cutoff could be at a lower energy,
which would increase the non-thermal energy density, too. If the low energy
electrons are in-situ accelerated the energy density in turbulence has to be
close to equipartition with the thermal energy density, as well. Although
this is not ruled out, it looks very unlikely.  The required amount of
relativistic electrons could be lowered if
\begin{enumerate}
\item[(a)]
important sources of optical photons were omitted. This could be e.g. a
population of intra-cluster stars, tidally stripped from galaxies
(Gregg \& West 1998). Their light is not included in the
luminosity function used. But the total emission of such stars is
limited to be 25--30\% of the total light from the cluster (Melnick et
al. 1977, Thuan \& Kormendy 1977). Thus the required relativistic
electron energy density could be lowered only by up to 25--30\%.
\item [(b)]
One might ask, if also IR- or UV-radiation fields of galaxies might give
important contributions to the EUV flux. For an electron distribution with a
single differential number index $\ale = 2.5$, the contribution of a specific
target population of photons with number density $n_{\rm ph}$, and photon
energy $E_{\rm ph}$ to the (fixed) observation band scales linearly with
$E_{\rm ph}^{(\ale-1)/2}\,n_{\rm ph} \sim E_{\rm ph}^{0.75}\,n_{\rm
ph}$. Since the IR-, and UV- radiation fields have much lower energy
densities than the optical photons, their contribution can be neglected.
\item[(c)]
The required energy can be strongly lowered if the correlation between
electron density and starlight photon density is not only valid on a global
cluster scale, as used in the above model, but also holds on the scale of
individual galaxies, as should be demonstrated in the scenario described in
the {Appendix}.
\end{enumerate}

\subsection{Difficulties of starlight-IC-HEX}

Scattering of starlight into the HEX region in order to explain the observed
excess needs a normalization constant of $C_{\rm e,o} = 2\cdot 10^{-4}\,{\rm
cm^{-3}}\,h_{50}$ for $\ale =2.5$ and the $\beta$-model assumed above. Since
the energy range of these electrons (84--150 MeV) overlaps with that of
CMB-IC-EUV scattering electrons (140--350 MeV) both populations have to
match. But the produced EUV emission would exceed the observed EUV excess by
a factor of roughly 20. Unless starlight-IC enhancement (see
{Appendix}) by a factor of 20 exists -- which does not seem to be very
likely -- the starlight-IC model as the mechanism producing the HEX excess is
ruled out. 

\section{HEX bremsstrahlung \label{sec:Brems}}

The HEX-excesses is emission above what is expected for a thermal electron
population. But the assumption of an exact thermal equilibrium might be
questioned, since turbulence is present within clusters due to the stirring
motion of galaxies, infalling subclumps, injection of radio lobes into the
ICM, and streaming motion of the ICM itself, e.g. in cooling flows. This
turbulence drags electrons out of the thermal tail, whenever they get into
resonance with plasma waves, and accelerate them to higher energies. This
leads to a modification of the distribution function so that the high energy
cutoff of a thermal distribution is replaced by some power-law like region.

Further, since it is known from the radio halo that a relativistic population
of electrons exists, there could be a spectral connection between
relativistic and thermal populations: the electrons cooling from relativistic
energies into the thermal population can add a power-law tail to the
thermal distribution.

Therefore it is reasonable to explain the measured HEX emission with a
modified thermal distribution function of the electrons. The EUV emission,
which is energetically below the thermal bulge, can of course not be
explained in this way.

The thermal electron distribution with temperature $k_{\rm B}T = 8.21$
keV (Hughes et al. 1993) can be written as a trans-relativistic
Maxwell-Boltzmann distribution in the dimensionless momentum $p=
P_{\rm e}/(\me c)$
\begin{equation}
f_{\rm e,th}(r, p)\, dp = \frac{n_{\rm e,th}(r)\, \beta_{\rm
th}}{K_2(\beta_{\rm th})}\, \, p^2 \,\exp (-\beta_{\rm
th}\,\sqrt{1+p^2})\, dp,
\end{equation}
with $K_\nu$ denoting the modified Bessel function of the second kind
(Abramowitz \& Stegun 1965\nocite{abramowitz65}), introduced here for
proper normalization, $\beta_{\rm th} = \me\, c^2/k_{\rm B}T$ the
normalized thermal beta parameter, and
\begin{equation}
n_{\rm e,th}(r) = n_{\rm e,o}\, (1+(r/r_{\rm c})^2)^{-\frac{3}{2}
\beta}
\end{equation}
 the number density of thermal electrons, parametrized with the
usual $\beta$-profile: $n_{\rm e,o} = 2.89\cdot 10^{-3}\, {\rm
cm^{-3}}\,\h^{1/2}$, $\beta = 0.75$, and $r_{\rm c} = 400\, {\rm
kpc}\, \h^{-1}$ (Briel et al. 1992\nocite{briel92}). We add a
power-law tail to this by writing 
\begin{equation}
f_{\rm e}(r,p)\, dp = \left\{
\begin{array}{ll}
f_{\rm e,th}(r, p); &\;\;\; p\leq p_*\\
C_{\rm e}(r) p^{-\ale}; & \;\;\; p>p_*
\end{array}
\right\}\, dp,
\end{equation}
where the normalization parameter is determined from the condition of a
smooth matching of the non- and thermal part to be $C_{\rm e}(r) =
p_*^{\ale} \, f_{\rm e,th}(r, p_*)$. 

Assuming for illustration a number index of $\ale = 2.5$, and
choosing $p_* = 1/2$, which corresponds to a kinetic energy of $E_*
=60$ keV, thus far above the typical thermal energy of 8 keV, gives
$C_{\rm e}(r) = 1.08\cdot 10^{-2}\, n_{\rm e,th}(r)$, or $C_{\rm e}(0)
= 3.13\cdot 10^{-5}\,{\rm cm^{-3}}\, \h^{1/2}$. This implies, that the
total number in electrons is only increased by less than 1.8~\%, even
if the power law is extended to infinity. The kinetic energy density
increases by less than 80~\%, or by only 12~\% if a higher cutoff is
introduced at $p=1$, which corresponds to a kinetic energy of 210
keV, sufficiently high in order to produce the observed emission up to
80 keV.

\begin{figure}
\psfig{figure=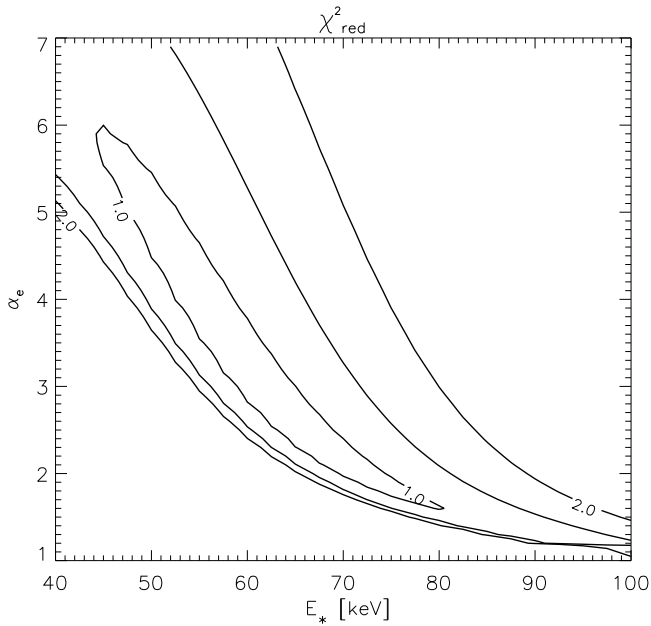,width=0.5\textwidth,angle=0}

\caption[]{\label{fig:chi} $\chi^2_{\rm red}$ for the bremsstrahlungs
model described in the text plotted for $E_*$ the matching energy of the
power-law and thermal electrons, and $\alpha$ the spectral index of the
former.  The OSSE and Beppo-Sax data is used simultaneously. }
\end{figure}

The bremsstrahlung emissivity spectrum is calculated via
\begin{equation}
q_{\rm X}(r,k)\, dk = n_{\rm e}(r)\, \int_0^\infty \!\!\!\! dp\, f_{\rm
e}(r,p)\, v(p)\,\sigma_{\rm X}(p,k)\, dk,
\end{equation}
where $k$ is the photon momentum in units of $\me\,c$, $v(p)$ the electron's
velocity, and $\sigma_{\rm X}(p,k)\,dk $ is the bremsstrahlung
cross-section. We use the transrelativistic interpolation formula, and the
Elwert factor given in Haug (1997) for a hydrogen plasma. Integrating this
over the emission volume within $1^\circ$ radius (or using the central
emissivity times an emission weighted volume of $0.36\, {\rm
Mpc^3}\,\h^{-3}$), and redshifting the spectrum, gives the observed flux,
which agrees well with the observed one, as can be seen in
Fig.~\ref{fig:HEX}.

The above assumed parameter $\alpha = 2.5$ and $p_* =1/2$
(corresponding to $E_* = 60$ keV) reproduce the Beppo-SAX and OSSE
data with a $\chi^2_{\rm red}$ of 1.6. A systematic $\chi^2_{\rm red}$
estimate is shown in Fig.~\ref{fig:chi}.  The absolute minimum is at
$\alpha = 2.9$, and $E_*= 62.5$ keV with $\chi^2_{\rm red} = 0.86$.
The length of the $\chi^2_{\rm red}$ valley in Fig.~\ref{fig:chi}
indicates that the necessary modification of the thermal distribution
is not strongly constrained.  Thus a very steep power-law tail, which
has a low energy content, explains the HEX excess well. Such a
spectrum can be expected, if in-situ acceleration produces this tail
by accelerating thermal electrons against the fast Coulomb cooling,
which has a cooling time of $t_{\rm C} \approx 10^5$ yr for 60 - 100
keV electrons (Spitzer 1968). Also models with a high energy cutoff
anywhere above 100 keV in the power-law tail give satisfactory
$\chi^2_{\rm red}$.

\section{Discussion\label{sec:disc}}
\subsection{EUV excess}
We discussed two inverse Compton processes, which could explain the
power law EUV excess of clusters of galaxies at the example of the
Coma cluster. 

The first one is IC scattering of the CMB. This requires a sharp cutoff in
the electron distribution above the relevant energy range of 140--350 MeV,
otherwise the electrons above this range overproduce low frequency radio
emission. In-situ acceleration produces an exponential cutoff, too soft to
hide this electron component. We propose that the relativistic electrons were
accelerated by a shock wave during a merger event of Coma, up to 2 Gyr
ago. Synchrotron and IC cooling produces a sharp cutoff in the energy
spectrum. In order to allow electrons to be still within the EUV producing
energy range after a cooling time of 2 Gyr, regions of lower than average
magnetic fields are necessary within the ICM, which is a natural
assumption. But the narrow spatial emission profile of the EUV excess is not
explained by this model. Maybe detailed simulation of particle acceleration
during a merger event can show that the resulting accelerated electron
profile is indeed as narrow as required.

The second model is IC scattering of starlight photons by electrons in
the energy range of 4--10 MeV. Since the target photon distribution is
peaked at the center of the cluster, the profile of the scattered
photons is narrower than that of the electrons. In order to
establish such an electron population, which would have a pressure
similar to that of the thermal background gas (but see
{Appendix} for the possibility of a much lower pressure),
confining magnetic fields have to separate these electrons from the dense
thermal background gas in order to prohibit the strong Coulomb losses. This
may be the case within remnant radio plasma from early epochs of quasar
activity, which did not mix with the gas, and which can form a second, nearly
invisible, non-thermal component of the ICM. Future high resolution EUV
observations may be able to resolve the possible enhanced IC emission near
the giant elliptical galaxies in Coma as it would be expected in the
starlight IC model. Since the starlight-IC model does not require a sharp
cutoff above the IC emitting electrons, IC flux of higher energy than the
thermal X-ray emission could be expected.  A high non-thermal pressure in
galaxy clusters was suggested to be a possible explanation of the discrepancy
between hydrostatic and lensing masses of clusters (Loeb \& Mao
1994\nocite{loeb94}, Miralda-Escud\'{e} \& Babul 1995, En{\ss}lin et
al. 1997)

\subsection{HEX excess}

Both IC processes discussed above lead to unsatisfactory implications for the
ICM, if they were considered as the explanation of the HEX excess:

IC scattering of CMB photons requires the ICM to be mainly unmagnetized ($B
\ll \mu$G), in order to avoid an overproduction of synchrotron emission by
the scattering electrons. But in order to be in agreement with current
measurements of ICM Faraday rotation, regions with strong field strength ($B
\gg\mu$G), but without a significant population of 3--5 GeV electrons have to
be present. This spatially very different electron content could be due to
spatially differentiated cooling, but only if no intermediate magnetized
regions ($B\approx \mu$G) exist in the ICM, otherwise the radio emission
would be too strong. Further, the injection of the electrons had to be a few
0.1 Gyr ago, not more or less, since otherwise the strong field region still
contains radio luminous electrons, or the unmagnetized regions lost their 3-5
GeV electrons due to IC-cooling.

The starlight IC model for the HEX excess requires a high electron number
density in the energy range of 84-150 MeV, which is too close to the energies
necessary to scatter the CMB photons into the EUV band as that an
overproduction of EUV flux can avoided. Only if the enhancement factor for
starlight-IC introduced in the {Appendix} is higher than 20 this
discrepancy can be solved, which seems to be very extreme.

We propose a third explanation of the HEX excess, which is not in conflict
with observations at other wave bands. Due to galactic motions, infall of
subclusters, convective movement of the gas, etc., turbulence is present and
therefore in-situ acceleration is expected to drag electrons out of the
thermal distribution and to accelerate them. This might lead to a power-law
like high energy tail of the distribution function above $\approx 50$
keV. The HEX excess is satisfactorily explained by the bremsstrahlung of
these electrons.

\section{Conclusion}

We conclude, that the EUV excess can be explained either with the CMB-IC
model, which has the advantage of a low energy density in the required
relativistic electrons, or with starlight-IC, which requires a high energy
density, but explains naturally the observed narrow emission profile.  The
latter model would be supported if future observations with high spatial
resolution discover a correlation between the location of luminous galaxies
and the EUV flux. Both models favor a strongly inhomogeneous magnetized
ICM. The electrons of 140--350 MeV required in the CMB-IC model should
have aged after injection for up to 2 Gyrs within the inhomogeneous fields,
in order to be consistent with radio constraints. This favors as an origin of
these electrons shock acceleration during a past cluster merger. The origin
of the 4.4-11 MeV electrons in the starlight-IC model is expected to be the
radio plasma outflow of earlier radio galaxies.  The CMB- and the
Starlight-IC model do not exclude each other.  Therefore a combined
contribution is possible.

The HEX excess is most likely bremsstrahlung of a supra-thermal tail
above 50 keV of the electron energy distribution, tracing acceleration
and cooling processes of transrelativistic electrons.

Future observations, which extend the observed spectral range of these
non-thermal excess emissions, and improve the spatial resolution, would give
further important insight into the complex trans- and ultra-relativistic
processes within the ICM, and its inhomogeneous structure.

\acknowledgement
We acknowledge very useful discussions with Thomas Bergh\"ofer, Christian
Zier, and Phil Kronberg. We thank Stuart Bowyer, Thomas Bergh\"ofer, Roberto
Fusco-Femiano, and his co-workers for giving us their data. This manuscript
was considerably improved due to comments by an anonymous referee. TAE
acknowledges support by the {\it Studienstiftung} and the MPIfR.

\begin{appendix}
\section{\label{sec:enhance}Enhanced starlight-IC}

The transition between ICM and ISM of individual elliptical galaxies may be
smooth in the sense that a galaxy drags the local ICM into its potential, and
compresses it thereby adiabatically. This could explain the observed presence
of X-ray peaks located at the positions of cD galaxies of clusters rather
than at the center of the cluster wide X-ray emission (Lazzati
\& Chincarini 1998). We model the background gas around a cluster galaxy by
\begin{equation}
X(r) = \frac{n_{\rm e,gal}(r)}{n_{\rm e,cl}} = 1 +
\frac{X_o-1}{1+r^2/r_{X}^2},
\end{equation}
where $n_{\rm e,cl}$ is the background density of the ICM. $X_o$, which we
assume in our example to be $X_o =7$, is the maximal compression of the ICM
density the galaxy achieves in its potential on a scale of $r_{X}$. The
radius from the center of the galaxy is $r$.

First, we estimate the change of conventional observables due to this
compression.  The temperature increases by a factor of $X^{2/3}(r)$. The
enhancement in X-ray emission is $X^{7/3}(r)$, since $L_{\rm X}\sim n_{\rm
e}^2\,T^{1/2}$. Averaging this over a typical volume a galaxy occupies within
the cluster, which we tentatively assume to be a sphere with radius $10\,
r_{X}$, one gets an increased emissivity by a factor of 1.5 over the same
volume filled by background gas. An estimate of the background gas
properties, assuming a homogeneous medium and using the observed X-ray data,
should result in a (emission weighted) gas density and temperature of
$1.16\,n_{\rm e,cl}$ and $1.21\, T_{\rm cl}$, only moderately different from
the background values $n_{\rm e,cl}$, $T_{\rm cl}$.
 
The population of relativistic particles is also adiabatically
compressed. Pressure equilibrium between the thermal gas and relativistic
electrons gives a compression factor of $Y = X^{5/4}$, due to the ratio
of adiabatic index of thermal gas and relativistic electrons (and also
magnetic fields) of $\frac{5}{3}/\frac{4}{3}$. An electron increases its
momentum during compression by a factor $Y^{1/3}$, so that a power law
distribution with differential number index $\ale$ gets an increased
normalization by a factor $Y^{(2+\ale)/3} = X^{5(2+\ale)/12}$.
The volume averaged intensity of IC scattering of a homogeneous photon
population around our model galaxy would be 1.34 times the background value,
for $\ale = 2.5$. But for photons emitted in the center, which have an
$r^{-2}$ density profile, and are therefore most abundant at the location of
highest electron densities, the volume averaged IC scattering for the
compressed relativistic electrons is 4.8 times the value the same photon
density profile would produce with a homogeneous electron profile.

\end{appendix}

\end{document}